\documentclass[epj,final]{svjour}

\usepackage{amssymb}
\usepackage{amsmath}
\usepackage{amsbsy}
\usepackage{bm}
\usepackage{color}
\usepackage{graphicx}

\newcommand{\bq}{\begin{eqnarray}}
\newcommand{\eq}{\end{eqnarray}}
\newcommand{\bqn}{\begin{eqnarray*}}
\newcommand{\eqn}{\end{eqnarray*}}
\newcommand{\rr}{\mathbf{r}}
\newcommand{\kk}{\mathbf{k}}
\newcommand{\RR}{\mathbf{R}}

\begin{document}

\title{Supercooled superfluids in Monte Carlo simulations}

\author{Riccardo Fantoni\thanks{\email{rfantoni@ts.infn.it}}}
\institute{
  Universit\`a di Trieste, Dipartimento di Fisica, strada
  Costiera 11, 34151 Grignano (Trieste), Italy
}


\abstract{
We perform path integral Monte Carlo simulations to study the
imaginary time dynamics of metastable supercooled superfluid states
and nearly superglassy states of a one component fluid of spinless
bosons square wells. Our study shows that the identity of the
particles and the exchange symmetry is crucial for the frustration
necessary to obtain metastable states in the quantum regime. Whereas
the simulation time has to be chosen to determine whether we are in a
metastable state or not, the imaginary time dynamics tells us if we
are or not close to an arrested glassy state.    
}

\maketitle

If a liquid can be cooled below its melting temperature $T_m$ without
the occurrence of crystallization, it is called a good glass former,
and when the temperature is less than $T_m$ the system is called
supercooled. The static and dynamical properties of such systems can
be studied over a large temperature range below $T_m$ and it is found
that their relaxation times increase very quickly by many decades if
the temperature is lowered. At a certain temperature the relaxation
time exceeds the timescale of the experiment and therefore the system
will fall out of equilibrium. It is this falling out of equilibrium
that is called the glass transition. At temperatures well below this
glass transition temperature no relaxation seems to take place any
longer, on any reasonable timescale, and it is customary to call this
material a glass. This transition temperature will in general depend
on the type of experiment, since its definition involves the timescale
of the experiment. Understanding the transition from a supercooled
liquid to a glass, or a disordered solid, is one of the major open
problems in condensed matter. 

In a liquid of number density $\rho$, made of mass $m$ particles,
moving in a $d-$dimensional space, the quantum effects will become
important when the temperature $T$ is comparable or smaller than the
degeneracy temperature $T_D=2\lambda\rho^{2/d}$, where
$\lambda=\hbar^2/2m$ and $\hbar$ is the reduced Planck constant. A
liquid such that $T_D>T_m$ is therefore likely to form a quantum
glass.  

At a temperature $T_\text{MCT}<T_m$ a kinetic glass transition towards
an arrested state is predicted by the Mode Coupling Theory (MCT)
\cite{Hansen-McDonald,Gotze}. Many of the qualitative predictions of
this theory have been confirmed in experiments and computer
simulations, and thus MCT can currently be regarded as the best
available theory of the dynamics of supercooled liquids. 

Our aim in this letter is to use Path Integral Monte Carlo (PIMC)
simulations \cite{Ceperley1995} to gain an understanding on the very
general question of the search for an arrested state when the
temperature approaches $T_\text{MCT}$. Since we are interested in an
universal property of glassy systems, our simulations are carried out
with a very simple and unrealistic model liquid, namely the
square-well bosons \cite{Fantoni2014a}. We will be working
at very low Temperatures $T\approx T_m<T_D$. We will find metastable
supercooled superfluid states and evidence for development towards a
superglass state \cite{Boninsegni2006c,Biroli2008,Hunt2009} which
should appear at even lower temperatures $T\approx T_\text{MCT}$.   

Using the terminology of Ref. \cite{Ceperley1995} we are then looking 
for local minima of the {\sl action} of the {\sl primitive
approximation}, up to thermal activation according to the Metropolis
algorithm \cite{MRTRT}. These may differ from the ones of the {\sl
  inter-action} due to quantum tunneling. In particular we will be
interested in how the identity of the particles and their exchange
permutation cycles which forms in a PIMC simulation frustrates the
development towards the global minimum of the action favoring the
formation of the metastable supercooled states \cite{Boninsegni2012}.

\label{sec:problem}
 
Consider a fluid (homogeneous and isotropic) of $N$ bosons in a
volume $V$ and density $\rho=N/V$  at a given absolute temperature 
$T=1/k_B\beta$, with $k_B$ Boltzmann constant, with a Hamiltonian
${\cal H}=-\lambda\sum_{i=1}^N\bm{\nabla}_i^2+\sum_{i<j\le N}\phi(|\rr_i-\rr_j|)$
symmetric under particle exchange, with $\lambda=\hbar^2/2m$, $m$ the
mass of the particles, and $\phi(|\rr_i-\rr_j|)$ the pair-potential of
interaction between particle $i$ at $\rr_i$ and particle $j$ at
$\rr_j$. The dynamic structure factor is defined as follows  
$
S(k,\omega)=\frac{1}{2\pi N}\int_{-\infty}^{\infty}dt\, e^{-i\omega t}
\langle\rho_{-\kk}(0)\rho_{\kk}(t)\rangle\\
=\int_{-\infty}^{\infty}dt\, e^{-i\omega t}F(k,t),
$
where $\rho(\rr)=\sum_{i=1}^N\delta(\rr-\rr_i)$ with
$\langle\rho(\rr)\rangle=\rho$, $\rho(\rr,t)=e^{i{\cal
    H}t/\hbar}\rho(\rr)e^{-i{\cal H}t/\hbar}$, $\rho_\kk(t)=\int
d\rr\, e^{i\kk\cdot\rr}\rho(\rr,t)$, and $\rho_\kk(0)=\rho_\kk$. 
Given an observable ${\cal O}$ we define the statistical average
as $\langle{\cal O}\rangle=\text{Tr}\left({\cal O}e^{-\beta {\cal
    H}}\right)/Z$ with $Z=\text{Tr}(e^{-\beta {\cal H}})$ the
partition function. 

We introduce the analytic continuation of
$F(k,t)=\int_{-\infty}^{\infty}d\omega\, e^{-\hbar\omega
  t}S(k,\omega)$ in imaginary time as follows 
\bq \label{fkt1}
F_k(t)&=&\frac{1}{NZ}\text{Tr}
\left(\rho_{-\kk}e^{-t{\cal H}}\rho_\kk e^{-(\beta-t){\cal H}}\right).
\eq
So that $F_k(0)=2\pi F(k,0)=\int_{-\infty}^\infty d\omega\,
S(k,\omega)=S(k)$ is the static structure factor such that
$\lim_{k\to\infty}S(k)=1$. 

Clearly we have that $F_k(t)=2\pi F(k,i\hbar t)$ is defined for $t\in
[0,\beta]$ being symmetric respect to $t=\beta/2$ since
$S(k,-\omega)=e^{-\beta\omega}S(k,\omega)$. 

The calculation of $F_k(t)$ of Eq. (\ref{fkt1}) becomes straightforward
in Path Integral Monte Carlo (PIMC) \cite{Ceperley1995} where it is
sufficient to average the product of $\rho_{-\kk}$ on the first
time-slice with $\rho_\kk$ at a time-slice a time $t$ later.

\label{sec:discussion}

The dynamic structure factor for the ideal Bose gas for particles of
spin $s$ at a temperature $T$ below the critical temperature
$k_BT_c=4\pi\lambda\{\rho/[(2s+1)\zeta(3/2)]\}^{2/3}$, where $\zeta$ is 
the Riemann zeta function, is given by Eq. (18) in Ref.
\cite{Baerwinkel1971},
where their $\lambda$ is our $\sqrt{4\pi\lambda\beta}$, the de Broglie
wave-length. 

In particular one finds
$
\left.\frac{dF_k(t)}{dt}\right|_{t=0}=
-\int_{-\infty}^\infty d\omega\,\hbar\omega S(k,\omega)=-\lambda k^2
$.

\begin{figure}[htbp]
\begin{center}
\includegraphics[width=8cm]{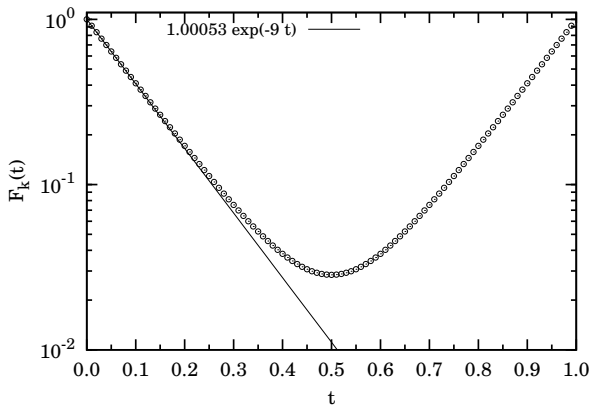}
\end{center}  
\caption{Behavior of $F_k(t)$, as a function of the imaginary time
  $t\in [0,\beta]$, for an ideal Bose gas below its critical
  temperature at $k=3$, $\lambda=1$, $k_BT=1$, $\rho=0.4$, and
  $s=0$. The critical temperature is $k_BT_c=3.597$. The points are
  the numerical results from Eqs. (\ref{fkt1}) and Eq. (18) of Ref.
  \cite{Baerwinkel1971}. On the ordinates axis we use a logarithmic
  scale.}   
\label{fig:fkt-ideal}
\end{figure}

In Fig. \ref{fig:fkt-ideal} we show how $F_k(t)$ is well approximated
by a pure exponential decay $S(k)e^{-\lambda k^2 t}$ for $t\in
[0,\beta/2]$. 
%

\label{sec:simulations}

We performed grand canonical PIMC with the ``worm'' algorithm
\cite{Boninsegni2006a} for a system of spin zero square-well bosons in
three spatial dimensions. As usual the path $\RR(t)$ is discretized in
imaginary time $t$ extending from $t=0$ to $t=\beta=n_\tau \tau$ with
a time-step $\tau$. It is made of $Nn_\tau$ {\sl beads} of coordinates
$\RR(t)=\{(x_i(t),y_i(t),z_i(t)) ~~~\forall i=1,\ldots,N\}$ at each
time-slice $t=t_j=j\tau$. The particles pair-potential is as follows
\bq
\phi(r)&=&\left\{\begin{array}{ll}
+\infty & r<\sigma\\
-\varepsilon      & \sigma\le r <\sigma(1+\Delta)\\
0       & \sigma(1+\Delta)\le r
\end{array}\right.
\eq
We choose $\varepsilon>0$ as the unit of energies and $\sigma$ as the
unit of lengths. We then introduce a reduced temperature
$T^*=k_BT/\varepsilon$ (with $\beta^*=1/T^*$), a reduced density 
$\rho^*=\rho\sigma^3$, and a reduced chemical potential
$\mu^*=\mu/\varepsilon$. When the mass $m$ of the bosons and/or the 
depth of their attractive well $\varepsilon$ are sufficiently large,
{\sl i.e.} $\lambda^*=\lambda/(\varepsilon\sigma^2) \ll 1$ we are in
the classical limit. The classical 
fluid has been studied originally by Vega {\sl et al.} \cite{Vega1992}
who found that the critical point of the gas-liquid coexistence moves
at lower temperatures and higher densities as $\Delta$ gets smaller. 
The quantum mechanical effects on the thermodynamic properties of
nearly classical liquids can be estimated by the de Boer quantum
delocalization parameter $\ell=\sqrt{2\lambda^*}$. \cite{Young1980}
The phase diagram of the system in the quantum regime, $T^*\lesssim
T^*_D=2\lambda^*(\rho^*)^{2/3}$, has recently been studied by us
\cite{Fantoni2014a} with our quantum Gibbs ensemble MC
algorithm \cite{Fantoni2014d}.

Unlike the work of Biroli {\sl et al.} \cite{Biroli2008} we will work
far away from the sticky limit \cite{Baxter1968} obtained by setting
the stickiness parameter ${\cal T}^{-1}=12e^{\beta\varepsilon}\Delta$
and taking the double limit $\varepsilon\to\infty$ and $\Delta\to 0$
at fixed ${\cal T}$. We could reach numerically such limit by taking
$\Delta$ small enough \cite{Fantoni13c,Fantoni13h,Fantoni15}. Instead
we will fix $\Delta=0.5$ in all cases as was done in the previous
analysis of Ref. \cite{Fantoni2014a}.

In the present letter we want to study the relaxation to zero of the
$F_k(t)$ in the quantum regime, so we must choose $\lambda^*\gg 0$ and
$T^*\lesssim T^*_D$. Choosing $\lambda^*=1$ we must choose a sufficiently
small temperature and a sufficiently high density. For $T^*=1$ we need
a reduced density $\rho^*\gtrsim (1/2)^{3/2}=0.35$. The maximum
reduced density allowed for our system is $\sqrt{2}=1.41$ for the
close packed configuration of the hard cores. The small
attraction between the particles will be responsible for a shift at
lower packing fractions, $\eta=\pi\rho\sigma^3/6$, of the melting
value for pure hard-sphere (which in the classical limit is
approximately $0.54$). 

In our PIMC we had to choose a discretization time-step,
$\tau^*=\beta^*/n_\tau$, for the imaginary time extending from
$t\varepsilon=0$ to $t\varepsilon=\beta^*$. We then chose
$n_\tau=100$ time-slices \cite{Ceperley1995}.  
The ``worm'' algorithm uses a menu of 9 different moves: advance,
recede, insert, remove, open, close, swap, wiggle, and
displace. Labeling each of these moves with $q=1,2,\ldots,9$ 
respectively, a single random attempt of any one of them with
probability $G_q=g_q/\sum_{q=1}^9g_q$ constitutes a MC step.
In our simulations we always chose $g_q=1$ for $q=1,2,\ldots,7,9$,
and $g_8=10$. For each move, except the displace one, a maximum number
of time-slices involved, $\overline{m}$, is also defined 
\cite{Boninsegni2006a} to control their acceptance ratios. We always
chose $\overline{m}_q=5$ for all $q$. For the displace move we chose a
displacement of the path of the order of $V^{1/3}/1000$. We always
chose the $C$ parameter defined in Ref. \cite{Boninsegni2006a} equal
to $0.1$. This value ensured an acceptance ratio for the $Z-$sector
\cite{Boninsegni2006a} lower but close to $1/2$ even if in the
simulations converging towards the solid state this increased passed
$1/2$.  

Our simulations were $5\times 10^4$ blocks long with one block made by
$100$ steps where we did not accumulate the averages and by $100$
steps where we did. This sets the simulation (experiment) time.

\label{sec:results}

We studied the model with $\Delta=0.5$ and $\lambda^*=1$ at $T^*=1$,
$V=100\sigma^3$, and $\mu^*=50,80,100$. Starting from
the empty box we reached a stable superfluid for $\mu^*=50
[\text{stable}],80 [\text{stable}]$ and a stable normal solid for
$\mu^*=100 [\text{stable}]$. Then we lowered the temperature at
$T^*=0.5$ and we studied the model with $\mu^*=80$. Now {\sl
  quenching} from the empty box we reached a metastable superfluid at
$\mu^*=80 [\text{metastable}]$ for the first $20000$ blocks which
later converged towards its stable normal solid state: $\mu^*=80
[\text{stable}]$. We then quenched from the empty box at a slightly
lower temperature $T^*=0.4$ keeping the chemical potential at
$\mu^*=80 [\text{metastable}]$ (which resulted in a slightly
higher density respect to the case at the higher temperature
$T^*=0.5$) and we observed that the system, instead of entering the
stable solid phase, stayed, for the whole length of our numerical
experiment, in a metastable supercooled superfluid state.

In Table \ref{tab:data} we report some properties of the simulated
system such as: the total energy per particle $e_\text{tot}$, the
kinetic energy per particle $e_\text{kin}$, the potential energy per
particle $e_\text{pot}$, the pressure $p$, the average number of
particles $\langle N\rangle$, the density $\rho=\langle N\rangle/V$,
and the superfluid fraction $\rho_s/\rho$, as calculated according to
Ref. \cite{Ceperley1995}. All the presented simulation were {\sl well
  converged} and the correlation simulation time $k_{\cal O}$ was
never bigger than $500$ blocks in any simulation for any property
${\cal O}$. The statistical error was as usual calculated as
$\sqrt{\sigma^2({\cal O})k_{\cal O}/N_{s}}$, where $\sigma^2(\cal O)$ is
the estimator of the variance of the random walk and $N_s$ the number
of MC steps.
\begin{table*}[htb]
\caption{Reduced properties of the simulated system with $\Delta=0.5$ at
  $\lambda^*=1, V=100\sigma^3$ and different $\mu^*$. For
  the simulation at $T^*=0.5,\mu^*=80 [\text{stable}]$ we
  considered the first $20000$ blocks as equilibration time and they
  were therefore discarded from the averaging. In all the other cases
  the equilibration time was taken equal to $1000$ blocks, {\sl i.e.}
  the ones necessary to bring the system from the empty box to the
  equilibrium number of particles.} 
\begin{tabular}{|ll|lllllll|}
\hline
$T^*$ & $\mu^*$ & $e_\text{tot}/\varepsilon$ &
  $e_\text{kin}/\varepsilon$ & $e_\text{pot}/\varepsilon$ &
  $p\sigma^3/\varepsilon$ & $\langle N\rangle$ & $\rho\sigma^3$ &
  $\rho_s/\rho$\\   
\hline
1.0&50 [\text{stable}]& 12.81(6) & 17.70(7) & -1.889(6) & 3.33(2)  &
33.92(7) & 0.3392(7) & 1.05(8)\\ 
1.0&80 [\text{stable}] & 19.20(7) & 21.94(8) & -2.734(8) & 6.20(3) &
42.41(8) & 0.4241(8) & 1.1(1)\\ 
1.0&100 [\text{stable}] & 24.12(6) & 27.46(7) & -3.335(7) & 8.75(3) &
47.79(6) & 0.4779(6) & 0.03(1)\\ 
0.5&80 [\text{stable}] & 17.029(8) & 20.325(8) & -3.297(3) & 6.504(3)
& 48 & 0.48 & 0.013(4)\\ 
0.4&80 [\text{metastable}] & 13.64(4) & 17.09(5) & -3.446(5) & 5.72(2)
& 50.19(5) & 0.5019(5) & 1.2(1)\\
0.4&90 [\text{metastable}] & 15.23(4) & 18.98(5)& -3.744(6) & 6.73(2)
& 53.16(5) & 0.5316(5) & 1.05(8)\\   
\hline
\end{tabular}
\label{tab:data}
\end{table*}

\begin{figure}[htbp]
\begin{center}
\includegraphics[width=8cm]{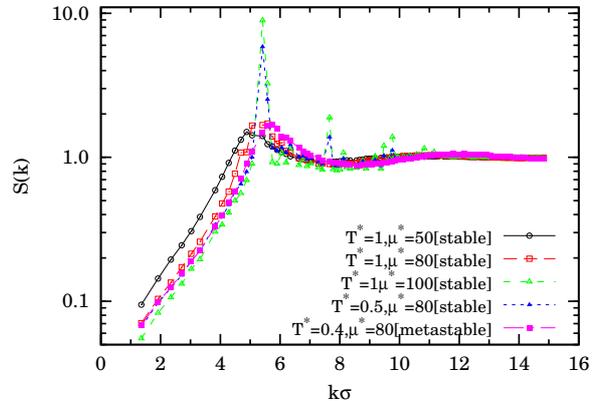}
\end{center}  
\caption{(Color online) Static structure factor $F_k(0)=S(k)$ at
  $\lambda^*=1$ and $T^*=1,\mu^*=50 [\text{stable}],80
  [\text{stable}],100 [\text{stable}]$ and $T^*=1/2,\mu^*=80
  [\text{stable}],80 [\text{metastable}]$. On the ordinates axis we use a
  logarithmic scale.}   
\label{fig:sk}
\end{figure}

\begin{figure}[htbp]
\begin{center}
\includegraphics[width=8cm]{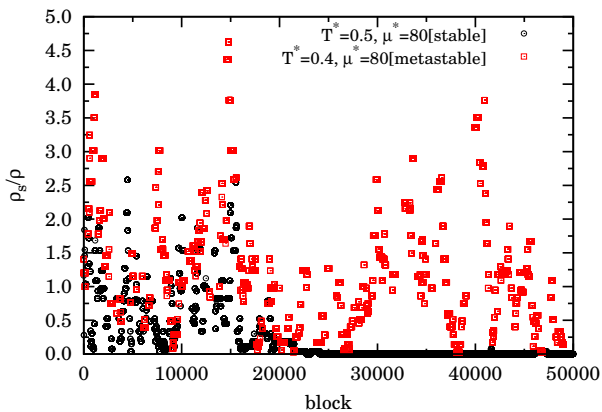}
\end{center}  
\caption{(Color online) Superfluid fraction at each PIMC block during
  the simulation at $\lambda^*=1$ and $T^*=0.5, \mu^*=80
  [\text{stable}]$ and $T^*=0.4, \mu^*=80 [\text{metastable}]$.}    
\label{fig:rhos}
\end{figure}

In Fig. \ref{fig:sk} we show the static structure factor of the
first five systems. This clearly shows how the $T^*=0.5,\mu^*=80
[\text{stable}]$ case is a solid state (the structure factor peak is
between $6$ and $7$) whereas the $T^*=0.4,\mu^*=80
[\text{metastable}]$ one is a fluid state (the structure factor peak
is here between $1.6$ and $1.8$). Note that in all cases the
simulation was $5\times 10^4$ blocks long and the acceptance ratio of
the $Z-$sector comparable. The difference between the two cases
immediately also appears by looking at the evolution of the superfluid
fraction during the progress of the simulations, as shown in
Fig. \ref{fig:rhos}. We clearly see how the $T^*=0.5,\mu^*=80
[\text{stable}]$ case has a transition from a superfluid state, before
block $20000$, to a normal solid, after. 
The behavior of $F_k(t)$ as a function of the imaginary time for some
chosen reciprocal wave-numbers around the first peak of the
correspondent static structure factor for the system with $T^*=0.4$
and $\mu^*=80$, which is a precursor of a superfluid glass, a superglass
\cite{Reichman2005,Boninsegni2006c,Biroli2008}, is such that we
observe exponential decays going below $10^{-2}$ for
$t\varepsilon>0.6$. Whereas for the systems in the solid state at
$T^*=1,\mu^*=100[\text{stable}]$ and $T^*=0.5,\mu^*=80[\text{stable}]$
we observe an almost constant value for $F_k(t)$ at the wave-number of
the first peak of the correspondent static structure factor and
exponentially decaying the other wave-numbers.  

\begin{figure}[htbp]
\begin{center}
\includegraphics[width=4.2cm]{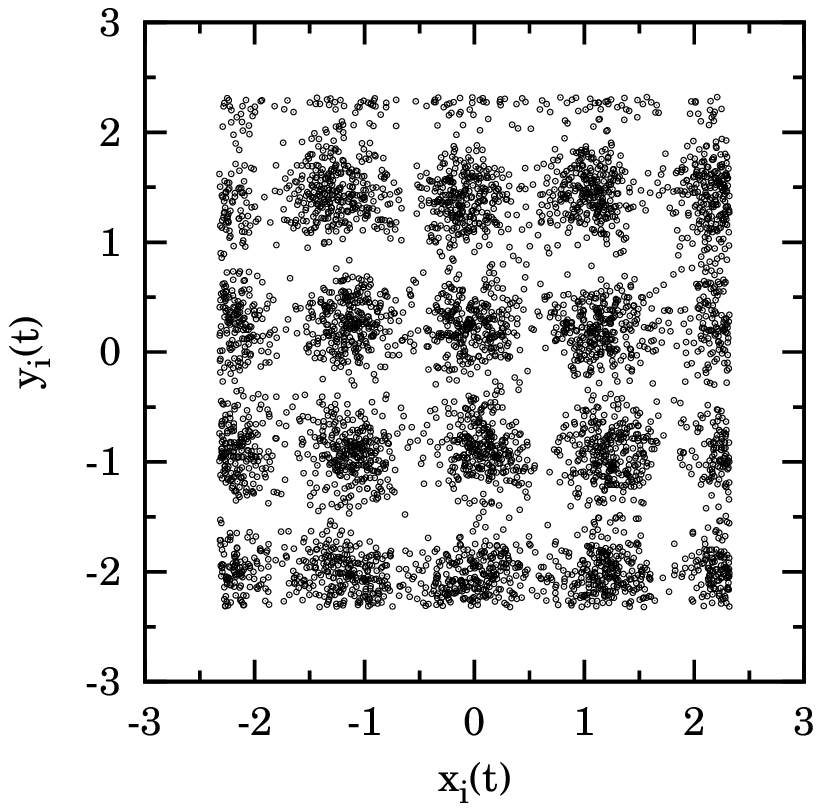}
\includegraphics[width=4.2cm]{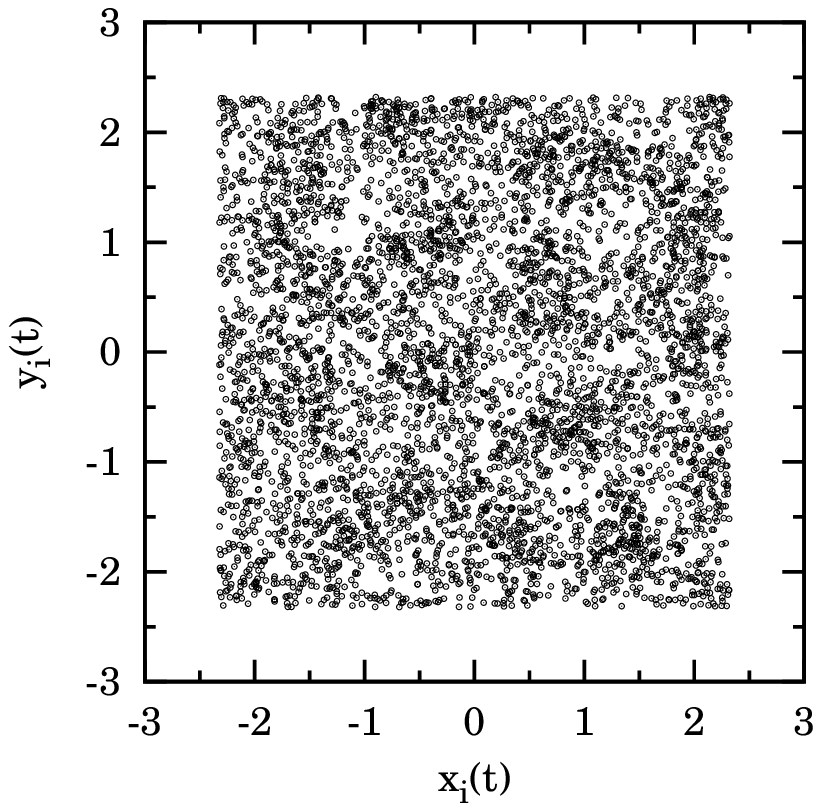}
\end{center}  
\caption{The $(x_i(t),y_i(t))$ particles positions at all
  time-slices at the end of the simulation with: $\lambda^*=1,
  T^*=0.5, \mu^*=80 [\text{stable}]$ (left panel) and $\lambda^*=1,
  T^*=0.4, \mu^*=80 [\text{metastable}]$ (right panel).} 
\label{fig:coord-l-80s}
\end{figure}

In Fig. \ref{fig:coord-l-80s} we show the $(x_i(t),y_i(t))$ particles
positions at all time-slices at the end of the simulation for the cases
$T^*=0.5,\mu^*=80 [\text{stable}]$ and $T^*=0.4,\mu^*=80
[\text{metastable}]$ respectively.  

Regarding the size effects we can say that the solid state we
observed has a triclinic lattice structure with a unit cell with
base vectors ${\bm a}=(0,0,a), {\bm  b}=(a,0,a/2), {\bm c}=(0,a,a/2)$
accommodating approximately $48$ particles. At $T^*=0.4$, a chemical
potential of $\mu^*=80$ is sufficient to reach approximately $50$
particles which could be adjusted in a different unit cell with the
same crystal structure. Thus we think that the size effects should not
be considered as responsible for the observed metastability.  

\label{sec:mode}

\begin{figure}[htbp]
\begin{center}
\includegraphics[width=8cm]{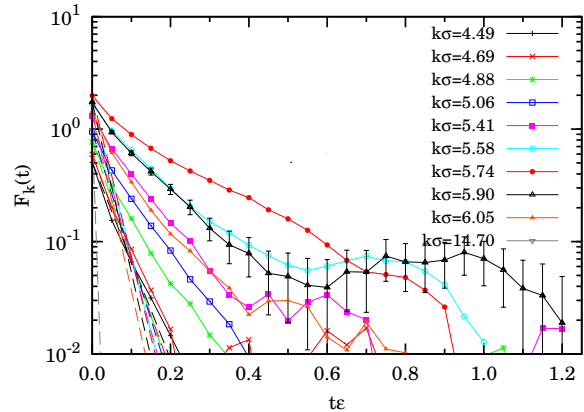}
\end{center}  
\caption{(Color online) The behavior of $F_k(t)$ as a function of the
  imaginary time for various values of $k$ around the first peak of
  the structure factor. We used $\lambda^*=1, T^*=0.4$, and $\mu^*=90
  [\text{metastable}]$. The dashed lines are the approximate
  ideal gas results. On the ordinates axis we use a logarithmic
  scale. For $k\sigma=5.90$ also the statistical errors are shown.} 
\label{fig:fkt-l-90m}
\end{figure}

In order to get closer to an arrested metastable state we restarted
from the the equilibrated supercooled superfluid configuration of
$T^*=0.4, \mu^*=80 [\text{metastable}]$ and increased $\mu^*$ by
$10$. This allowed us to reach another metastable supercooled
superfluid state closer to an arrested state where the $F_k(t)$, for
the $k$ around the first peak of the static structure factor at 
$2$, shows an  initial exponential decay followed by a plateau. This
is clearly shown in Fig. \ref{fig:fkt-l-90m} taken at the end of the
simulation and is in accord with the MCT predictions. In order to
observe the plateau it is essential the restarting or {\sl aging}
procedure. 
%

\label{sec:conclusions}

In conclusion, we proved, for the idealized model of spinless square
well bosons, that superfluidity is able to sustain metastability at low
temperature and high density. In order to define whether we are on a
metastable state we need to fix a simulation time interval much longer
than the correlation simulation time. We were able to maintain the
system in a metastable supercooled superfluid state for a rather long
simulation time. The metastable state may not be unique and there may
be many of those for a given set of thermodynamic conditions ({\sl e.g.}
$\mu,V,T$ in the grand canonical ensemble) all different from one
another depending from the kind of quench. The real (diffusive)  
dynamical (imaginary) time of the physical system can be used to
define the insurgence of an arrested glassy state through the aging
procedure, even if it is limited to the interval $[0,\beta/2]$.    

We should mention here that the simulation time for a classical
molecular dynamic and for a MC numerical experiment have
profoundly different meanings. The first one can be mapped into the
real dynamical time of the classical physical system whereas the
second one has nothing to do with it but is merely the number of
stochastic moves made to sample the configuration space of the system
within the Metropolis algorithm. In the quantum regime one has at his
disposal only simulations of the MC type but, as we showed,
the simulation time can give an indication of metastability. Whereas
the imaginary time real dynamics of the system tells us if we are
close to an arrested glassy state. 

We are presently implementing a better hard-core propagator
\cite{Cao1992} to substitute to the primitive approximation which
would allow us to use fewer time-slices. 

\bibliographystyle{unsrt}
\bibliography{qg}

\end{document}